# Petri-net modeling of B-cell receptor signaling pathways: A case study in CLL

**Gajendra Pratap Singh, *Madhuri Jha
*Mathematical Sciences and Interdisciplinary Research Lab, School of Computational and Integrative Sciences (SCIS)
Jawaharlal Nehru University, New Delhi-110067
**gajendra@mail.jnu.ac.in, *jhamadhuri81@gmail.com

**Abstract**

Immunology is the emerging research area which deals with the study of immune system in any living organism. It is modeled through various computational and mathematical models to deal with the problem facing while to boost the immune system of any organism or to fight with the infectious disease at the very initial stage. Such models are very important for a better understanding of the complex behavior of pathways inside the cells. The signaling pathways between the cells are complex and difficult to visualize in the immune system of human beings. So, it's important to study about the function of these cells separately. T-cells and B-cells are the important part of immune system and both have their own receptors and their different signaling pathways by which they deal with any antigens. In this paper we discuss about B-cell receptor (BCR) and its different signaling pathways downstream of the BCR. We designed a Petri-net model of the process of gathering antigens through B-cells independent of T-cell and the effect of that in immune system of the organism. We will also discuss the contribution of BCR in the selection of the precursor tumor cell in CLL (chronic lymphocytic leukemia).

## 1. Introduction

We are surrounded all the time the millions of bacteria or viruses and if it was not a well develop immune system then we would be infected by certain number of infections from time to time. The word *immunology* comes from the study that how an individual recovered from any infectious diseases and could be protected from the disease. The Latin term immunis, meaning "exempt", which means the process of protection from any infectious disease. So, it's important to build a healthy immune system so to be protected by several antigens. The Immune system of a human being is highly complex and a major challenge for scientists in the field of medicine, and pharmacology. The immune system is expanded in the whole body which involves certain different cells, tissues and organs [1, 2]. The two type of immune system is Innate or Acquired. The innate system recognizes the set of patterns that the antigens follow. On the other hand the acquired system recognizes several different types of antigens. The T and B-cells are the part of this acquired immune system. In particular B-cells participate in HUMORAL immunity and it is the only cells that synthesize immunoglobulin [3]. Immunoglobulin is the antigens that are secreted from the cells. The main components of this system are lymphatic system, lymphoid tissues, lymph nodes, thymus and spleen [4]. The main purpose is to study that how the immune system secretes antibodies that allow B-cells to bind with a specific antigen, to initiate the antibody response. Further BCR signaling constitutes different downstream pathways which act differently in different diseases [5]. The mature B-lymphocytes has an important role in identifying the proper precursor in tumor cell. CLL is the most common leukemia in the western countries. This CLL don't have strong responses to immune therapies so the advances have been made in chemotherapies and how to select the proper antibodies to prevent it on the initial stage [6]. So to get clear information it is important to model the system with a discrete mathematical modeling.

## 2. Motivation and Plan

Considering the importance of immune system it is better to study the function step by step. Biological networks these days are modeled with several computational and mathematical models such as differential equation, process algebra [7] and certain other field of science. Petri net graph are also one of the mathematical model which in these days are most in use to model several biological system [8, 9]. Petri net act as a tool which provides graphical and mathematical view of a system. Due to its dynamic discrete behavior it can show the interrelationship between the states involving in any system. Several high level Petri net tools [10,11] are also available to model the concurrent and synchronized system which makes Petri net a novel model in several field of science. In this paper we plan to model the B-cell receptor signaling pathways and their behavior in different deficiencies. We have also inspired to model the effect of BCR in CLL as it is taking one of the downstream pathways from the four pathways of BCR [12].

## 3. Biological background

### 3.1 B-cell receptor

**B-cell**, B-lymphocytes are a part of white blood cell originated and matured in the bone marrow. It is an important part of the Immune System. B-cell has a type of receptors on its cell surface which binds with the antigens and then it proliferates. B-cell activation can be T-cell dependent or T-cell independent. 95% of antigens are T-dependent but those which are T-independent are acting differently. In T-dependent antigens the immune system responses much stronger than T-independent as in that case memory cells are generated, induction of isotype switching occurs, and affinity maturation generated. The parts of the B-cell receptor (BCR) are shown in the Figure 1, which consists of antigen binding site, different regions and chains [13,14].

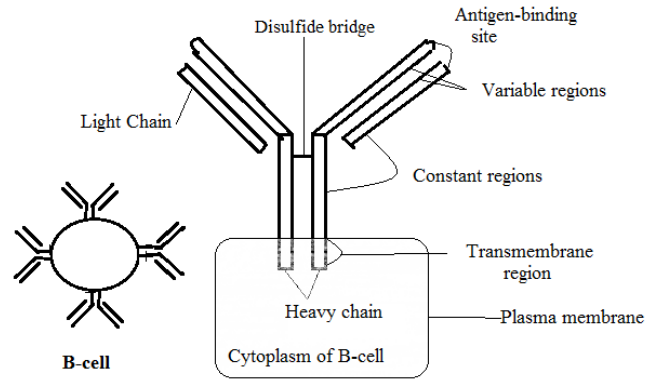

Figure 1: B-cell receptor (BCR)

## 3.2 BCR Signaling pathways

In this paper we specifically designed the model for B-cell activation independent to T-cell. B-cell generates plasma cell which secretes antibodies which protect the human body against the infections by binding the viruses and microbial toxins [15]. The magnitude and duration of BCR signaling are affected by the MAPK pathways, the P13K pathways, NFAT pathways, NF-$\alpha\beta$ pathways, and internalization of the BCR [15]. In general, B cells are often activated by antigen-presenting cells that capture antigens and display them on their cell surface. Activation of B cells by such membrane-associated antigens requires BCR-induced cytoskeletal reorganization [16]. Figure: 2, shows the different downstream pathway followed in BCR signaling and also the enzymes responsible for the functioning of BCR in immune system

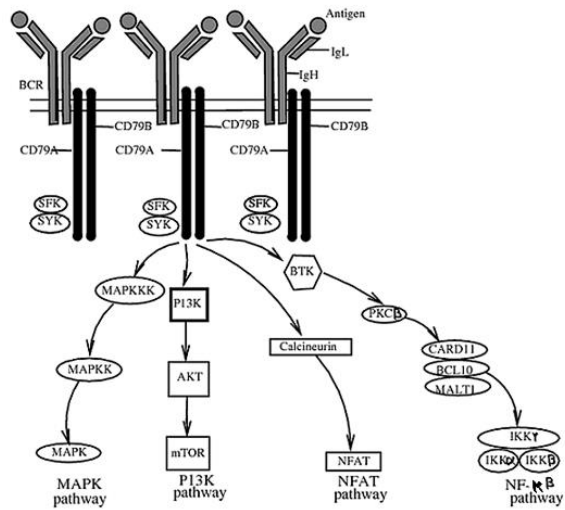

Figure 2: B-cell Signalling pathways

## 4. Methods

### 4.1 Petri net

Petri net is a special type of directed graph that can be used to study the dynamics of the modeled system. Petri net is 5-tuple set, consisting of a finite set of places represented as a circle, a finite set of transitions represented as a rectangle, positive incidence function that denotes the number of arc from any transition to place, negative incidence function that denotes the number of arc from any place to transition and the initial marking which represents the number of tokens on each places. Tokens are represented as black dots. Mathematically Petri net is represented as:

$$PN = (P, T, I^+, I^-, \mu^0)$$

Any transition fires at the given marking if and only if the outgoing arcs from all the places is less than and equal to the present marking of that place. After firing the tokens are moved from the input places and deposited on the output places accordingly. It changes the marking and a new marking obtained for the places associated with that transition. All these marking collectively form a reachability graph of that Petri net. This graph shows the reachable state from other states, which helps to obtained the relation between one state to another state. Several properties of Petri net is very beneficiary to draws some useful results of the modeled system [17, 18, 19].

### 4.2 Properties of Petri net

*State machine*: The Petri net in which each transition is strictly having one input and one output place.
*Marked graph:* A Petri net is a marked graph if each place is input for exactly one transition and also output for exactly one transition.
*Bounded*: A Petri net is bounded if the number of tokens on each place never be infinite, i.e., number of tokens remains finite.
*Free choice net*: It is an ordinary Petri net such that every arc from a place is either a unique outgoing or a unique incoming arc to a transition.
*Extended free choice net:* Any Petri net is extended free choice net if and only if for all transition there exists an another transition from the same net whose pre places set have at least one common place.
*Liveness and Reachability:* A Petri net is live if every transition of the net is live. Any transition is live in a marking $\mu$ if for each marking $\mu'$ there exists a marking sequence where that transition is enabled in marking $\mu$. If any of the transition is not live then that transition is known as dead transition and hence the Petri net is deadlock.
*Safeness*: A Petri net is safe if the number of tokens on any place never exceeds one.
*Simple net:* A Petri net is a simple net if each transition has at most one input place which is shared with another transition.
All these properties of Petri net are very helpful in studying any system [20, 21, 22].
Some basic structures which is not allowed to be a marked graph, free choice net and simple net are shown in Figure 3.

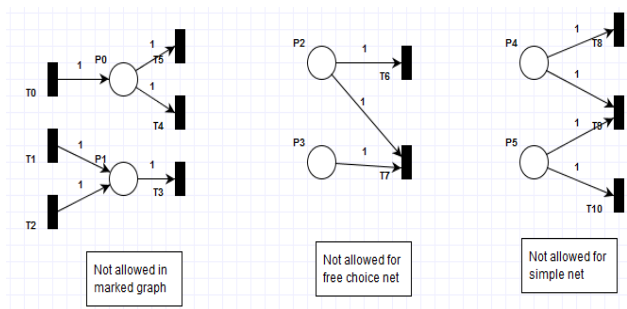

Figure 3: Forbidden structures

### 5. Modeling and Analysis

While modeling any system we first analyze the key elements, events and their relationships within the pathways. These steps help us to model the places and transition accordingly. Modeling the B-cell receptor signaling pathways we represents the molecular activities or stable compounds as places while the metabolic enzymes or responses (binding, proliferation etc) taken as transitions and the relationship between the activities with the corresponding enzymes are shown by the arcs. The tokens on the places shows the availability of the corresponding compound and flow of tokens means the use of one compound to combine with some enzymes to produce another compound. The Petri net model of B-cell receptor signalling pathways is shown in Figure 3.

### 5.1 Petri net model of B-cell receptor signaling pathways

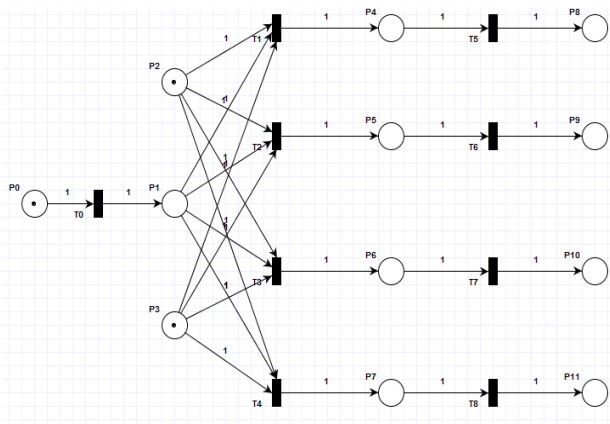

Figure 4: Petri net modeling of B-cell signaling pathway

| Places | Description | Transitions | Description |
|---|---|---|---|
| $p_0$ | Antigens | $t_0$ | Binding with IGL and IGH in BCR |
| $p_1$ | BCR | $t_1$ | Tyrosine kinase activity to generate MAPKKK |
| $p_2$ | CD79A+CD79B | $t_2$ | Tyrosine kinase activity to generate P13K |
| $p_3$ | SFK+SYK | $t_3$ | Tyrosine kinase activity and $Ca^{+2}$ mobilization |
| $p_4$ | MAKKK | $t_4$ | Tyrosine kinase activity to generate BTK |
| $p_5$ | P13K | $t_5$ | Phosphorylation |
| $p_6$ | Calcineurin | $t_6$ | Activation of IGF1 |
| $p_7$ | BTK | $t_7$ | Dephosphorylation and entry of NFATc protein |
| $p_8$ | Activation of MAPK pathway | $t_8$ | Proliferation and differentiation |
| $p_9$ | Activation of P13K pathway | | |
| $p_{10}$ | Activation of NFAT pathway | | |
| $p_{11}$ | Activation of NF-$\alpha\beta$ pathway | | |

### 5.2 Analysis of the model

To check the validity of model shown in Figure 4, we have taken one token in $p_0$ which means the availability of antigens inside the body and is ready to bind with b-cell receptor in the presence of IGH and IGR through the transition $t_0$. Firing of $t_0$ will implies the token shift to $p_1$ which is BCR which is responsible for the further steps to recover from the antigens effect. Now as the places $p_2$ and $p_3$ represents the availability of (CD79A + CD79B) and (SFK +SYK) enzymes which are already present in our body. So, we have assumed one token on both the places. In next step the transitions $t_1$, $t_2$, $t_3$, $t_4$ are enabled but only one will
fire depending on the antigen associated with BCR. These four transitions will decide which pathways will follow to reduce the adverse effect of the antigens. It all depends on tyrosine kinase activity that which pathway will lead to cure from the particular disorder. If $t_1$ will fire that means token shifts to place $p_4$ which is generation of MAPKKK protein. If $t_2$ will fire then the token shifts to $p_5$ which means the formation of P13K protein. Similarly if $t_3$ will fire token shifts to place $p_6$ which means the generation of calmodulin which is a well known calcium sensor protein which further activates to become phosphotase Calcineurin. If $t_4$ will fire then token shifts to $p_7$ which means the generation of BTK protein [23,24,25].

### 5.3 Validation

The model is validated from the online software PIPE (Platform Independent Petri Net Editor v4.3.0) [26].

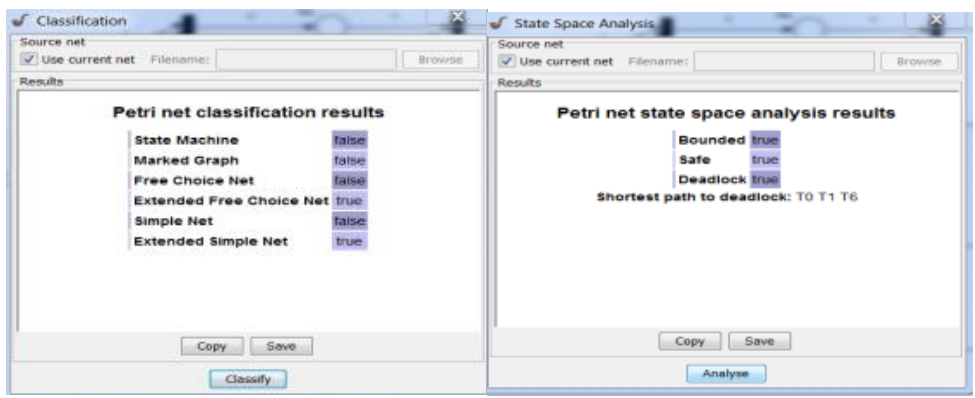

Figure 5: Property analysis with PIPE

From Figure 5 we can analyze that it is not a state machine as $t_1$, $t_2$, $t_3$, $t_4$ are the transitions having more than one input. As we know any enzyme present in the body is not responsible for a single proliferation rather it can involved in so many reactions. It is not a free choice and simple net but can be extended to follow both properties by modeling the enzymes differently with its specific properties. The system is bounded and safe means accumulation of any enzymes is not occurring. While studying about the different states in this model it is important to know about the path by which one state is reachable from another one . the Figure 6, shows the coverability graph of the model $S_6$, $S_7$, $S_8$, $S_9$ are the tangible states and rest are the vanishing states.

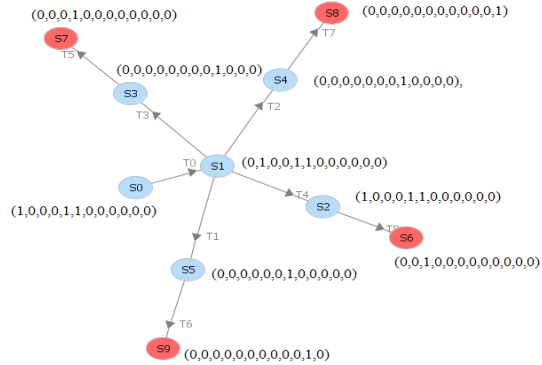

Figure 6: Coverability graph

In this coverability graph it has been validated that the four downstream pathways of BCR is independent from each other and originated from the same node *i.e.,* B-cell receptor.

## 6. Role of BCR in CLL

Chronic Leukemia is the cancer of leukocytes in which lots of partially developed white blood cells are present in blood which interfere the development and function of healthy blood cells and platelets. Two types of chronic leukemia can occur in the body either CML (chronic myeloid leukemia) [27] which affects granulocytes or CLL (chronic lymphocytic leukemia) [28] which affects lymphocytes especially b-cells. These premature cells which should die early are present in the blood and slow down the function of the mature cells and weaken the immune system of the body. These immature cells when come in the contact of BCR interfere with the cells BTK which stops the CLL cells to mature fully and interfering with other tyrosine kinases made them die slower before they divide. These premature cells contains proteins CD5, CD23 and CD19 on their surfaces which comes in the contact of B-cell receptors. BTK and P13K are the critical enzymes present in BCR which is responsible for the activation of AKT, NF-kβ pathways which promotes survival and proliferation of CLL cells [29].

## 7. Final summary and scope

As the researches in Immunology are becoming trends in the field of science and medicines so it is highly in need to model the system in different ways to study about the function and parts of the system properly. In this paper we have discussed the function of b-cell receptor when defending from the antigens coming from outside and attacking the human body. The main step is the one when it is being decided that which path should follow to cure the particular infection. The four pathway MAPK pathway, P13K pathways, NFAT pathways, and NF-αβ pathway works differently in different diseases.
For example the (P13K –AKT-mTOR) pathway helps to cure the breast cancer in early steps with the help of Phosphoinositide 3-kinase (P13K) inhibitor [30]. NFAT pathway is helpful to cure the progressive tumors in the body, where drugs are designed with excessive $Ca^{2+}$ signalling [31]. Phosphorylation of a tyrosine and threonine activates MAP kinase pathway which can suppress the growth of colon tumors. Similarly NF-αβ activation contributes to degrade various inflammatory diseases [32,33]. One can model these pathway differently to study their function in curing diseases. Petri net modeling helps to know the system in a proper way as we can predicts the liveness , boundedness, finding the deadlock, and the reachable states of the system.

**Acknowledgement**
This work is supported by DST purse and UPOE-2,(Id No: 257) and a project under DST-SERB ( Id no ECR/2017/003480/PMS).

**Abbreviations :** CD79A, CD97B: hetrodimers important in signaling component, SYK: spleen tyrosine kinase, SFK: SRC family kinase, MAPK: microtubule associated protein kinase, P13K: phosphoinositide 3-kinase, AKT(PKB): protein kinase B, mTOR: mammalian target of rapamycin, BTK: bruton tyrocine kinase, PKC: protein kinase, IKK: inhibitor of kappaB kinase, IGL: immunoglobulin lambda locus, IGH: immunoglobulin heavy locus, NF-αβ: protein complex controls transcription of DNA, IGF1: insulin –like growth factor 1.